\documentclass[11pt]{article}
\usepackage{moriond2,epsfig}
\usepackage{axodraw}
\usepackage{epsfig}
\usepackage{amsmath}
\usepackage{amssymb}
\usepackage{graphicx}
\usepackage{subfigure}
\usepackage{color}

\input epsf

\begin{document}
\vspace*{4cm}
\title{Loop Corrections to Scalar Quintessence Potentials}
\author{Michael Doran and  J\"org J\"ackel}

\address{Institut f\"ur Theoretische Physik der Universit\"at Heidelberg\\
  Philosophenweg 16, D-69120 Heidelberg, Germany}

\maketitle

\abstracts{
  The stability of scalar quintessence potentials under quantum
  fluctuations is investigated both for uncoupled models and models
  with a coupling to fermions. We find that uncoupled models are usually
  stable in the late universe. However, a coupling to fermions is
  severely restricted. We check whether a graviton induced
  \mbox{fermion-quintessence} coupling is compatible with this
  restriction.}


\newcommand{\omd}{\Omega _{\rm d}}
\newcommand{\sa}{\sigma _8}
\newcommand{\omdsf}{\bar{\Omega} _{\rm d} ^{\rm sf}}
\newcommand{\omdn}{\Omega _{\rm d}^0}
\newcommand{\omdeq}{\sqrt{1-\bar{\Omega}_{\rm d} (\aeq)}}
\newcommand{\omdroot}{\sqrt{1-\bar{\Omega}_{\rm d} (a)}}
\newcommand{\aeq}{a_{\rm eq}}
\newcommand{\adec}{a_{\rm dec}}
\newcommand{\atr}{a_{\rm tr}}
\newcommand{\wda}{w_{\rm d}}
\newcommand{\wdan}{w_{\rm d}^0}
\newcommand{\kmax}{k_{\rm max}}
\newcommand{\keq}{k_{\rm eq}}
\newcommand{\sla}[1]{\slash\!\!\!#1}
\newcommand{\slad}[1]{\slash\!\!\!\!#1}
\newcommand{\lferm}{\Lambda_{\rm ferm}}
\newcommand{\pmp} {\Phi} 
\newcommand{\pmpc}{\Phi_{\rm cl}} 
\newcommand{\gev}{\textrm{Gev}}
\newcommand{\eloop}{\textrm{1-loop}}
\newcommand{\vv}[2]{V_{\textrm{#1}}^{\textsc{#2}}}
\newcommand{\ww}[2]{W_{\textrm{#1}}^{\textsc{#2}}}
\newcommand{\fpi}{(2\pi)^{-4}}
\newcommand{\tpi}{(2\pi)^{-2}}
\newcommand{\mf}{m_{\rm f}(\Phi_{\rm cl})}
\newcommand{\mfn}{m_{\rm f}^0}
\newcommand{\mfnsq}{\left[\mfn\right]^2}
\newcommand{\mfsq}{\left[\mf\right]^2}
\newcommand{\mplank}{\textrm{M}_{\textrm{P}}}

\def\frac#1#2{\mathinner{#1\over#2}}

\section{Introduction}\label{introduction}
Observations indicate that dark energy constitutes a substantial
fraction of our Universe
\cite{Riess:1998cb,Perlmutter:1999np,Netterfield:2001yq,Lee:2001yp,twodf}.
The range of possible candidates
includes a cosmological constant and - more flexible - some
form of dark energy with time dependent equation of state, called
quintessence \cite{Caldwell:1998ii}.  Commonly, realizations of  quintessence scenarios
feature a light scalar field \cite{Peebles:1988ek,Ratra:1988rm,Wetterich:1988fm}.

The  evolution of the scalar field is usually treated at the
classical level. However, quantum fluctuations may alter the
classical quintessence potential\cite{Kolda:1998wq,Peccei:2000rz}.
In this talk which is based mainly on \cite{Doran:2002bc},
we have a look at one-loop contributions to the effective potential both from
quintessence and fermion fluctuations.
We will show that in the late universe, quintessence fluctuations
are harmless for most of the potentials used in the literature.
For inverse power laws and SUGRA inspired models, this has
been demonstrated in \cite{Brax:2000yv}. On the other hand, we will find that fermion
fluctuations severely restrict the
magnitude of a possible coupling of quintessence to fermionic dark
matter.

In Euclidean conventions, the action we use for the quintessence
field $\Phi$ and a fermionic species $\Psi$ to which it may couple
\cite{Wetterich:1988fk,Amendola:2000er,Bean:2001zm} is
\begin{multline}\label{equ::action}
S = \int d^4x\  \sqrt{g}\Bigg[\mplank^2R+\frac{1}{2}
\partial_{\mu} \Phi(x) \partial^{\mu} \Phi(x) + V(\Phi(x))  +
\bar\Psi(x)\left[i\ \slad{\nabla} + \gamma^5 m_{\rm f}(\Phi)
\right ]  \Psi(x)\Bigg],
\end{multline}
with $m_{\rm f}(\Phi)$ as a $\Phi$ dependent fermion mass. This $\Phi$
dependence (if existent in a model) determines the coupling
of the quintessence field to the fermions.
As long as one is not interested in quantum gravitational
effects, one may set  $\sqrt{g}=1$, $R=0$
and replace $\slad{\nabla}\rightarrow\sla{\partial}$
in  the  action \eqref{equ::action}.

By means of a saddle point expansion, we arrive at the
effective action $\Gamma[\pmpc]$ to one loop order of the quintessence
field.  The equation governing the dynamics of the quintessence field
is then determined by $\delta \Gamma[\pmpc] _{| \pmpc= \pmpc^\star} =
0$. When estimating the magnitude of the loop corrections, we will
assume that $\pmpc^\star$ is close to the solution of the classical
field equations: $\delta S = 0$.  Evaluating $\Gamma$ for constant fields,
we can factor out the space-time volume
$U$ from $\Gamma=UV$. This gives the effective potential
\begin{equation}\label{final}
\vv{\eloop}{}(\pmpc) = V(\pmpc) + \frac{\Lambda^2}{32\pi^2} V^{\prime\prime}(\pmpc) - \frac{\lferm^2 }{8\pi^2} \mfsq.
\end{equation}
Here, primes denote derivatives with respect to $\Phi$; $\Phi_{\rm cl}$ is the classical field value and $\Lambda$ and $\lferm$ are the
ultra violet cutoffs of scalar and fermion fluctuations. The second term
in Eq. \eqref{final}, is the leading order scalar loop.
We neglect graphs of the order  $(V^{\prime\prime}_{| \rm cl})^2$ and higher,
because $V$ and its derivatives
are of the order $10^{-120}$ (see section \ref{uncoupled}).  We have
also ignored $\Phi$-independent contributions, as these will not
influence the quintessence dynamics.
However, the $\Phi$-independent contributions add up to a cosmological
constant of the order $\Lambda^4 \approx {\mathcal{O}}(\mplank^4)$. This
is the old cosmological constant problem, common to most field
theories.  We hope that some symmetry\footnote{This symmetry must be unbroken or only very slightly broken.
E.g. SUSY is broken too badly to do this as has been pointed out in \cite{Kolda:1998wq}}
or a more fundamental theory will force it to vanish. The same symmetries or
theories could with the same right remove the loop contribution by
some cancelling mechanism. After all, this mechanism must be there,
for the observed cosmological constant is far less than the naively
calculated ${\mathcal{O}}(\mplank^4)$.

Besides, none of the potentials under investigation can be renormalized in the
strict sense.

There is also a loophole for all models that will be ruled out in
the following: The potential used in a given model could be the
full effective potential including all quantum fluctuations, down
to macroscopic scales. For coupled quintessence
models, this elegant argument is rather
problematic and the loophole shrinks to a point (see section
\ref{sec::coupling}).

We use units in which $\mplank =1$.

\section{Uncoupled quintessence}\label{uncoupled}
Let us calculate here only the simplest example, the pure
exponential $V(\pmpc)=A\exp(\lambda\pmpc)$ (for other potentials
see \cite{Doran:2002bc} and references
therein). Inserting into Eq. \eqref{final}
we find the one loop corrected potential:
\begin{equation}
\vv{\eloop}{ep} =  \vv{cl}{ep} \left \{ 1 +  \frac{1}{32\pi^2}  \Lambda^2 \lambda^2 \right\}.
\end{equation}
The potential is simply multiplied by a field independent constant. It is easy to see
that a rescaling $A\rightarrow A/(1+\frac{\Lambda^{2}}{32\pi^2}\lambda^{2})$ absorbs the
loop correction, leading to a stable potential up to order $V^{\prime\prime}_{\textrm{cl}}$.
The next to leading order $(V^{\prime\prime}_{\textrm{cl}})^2$ is non trivial but
very small since
\begin{equation}
\label{equ::small}
V^{\prime\prime}\sim V\sim H^{2}\sim 10^{-120}
\end{equation}
in the late universe. However, it is this term which spoils the strict renormalizability in four dimensions
for the pure exponential potential.\\
Other potentials are usually not completely form invariant even to lowest order. Nevertheless,
keeping in mind that $\frac{1}{32\pi^2}\approx 0.003$ and using reasonable cutoffs
$\Lambda\lesssim 1$ (remember we use units where $\mplank=1$) loop corrections are usually small
in the late universe. Therefore, most potentials are stable (for a hand picked exception see \cite{Doran:2002bc}).
\section{Coupled quintessence}\label{sec::coupling}
Various models featuring a coupling
of quintessence to some form of dark matter have been
proposed
 \cite{Wetterich:1995bg,Amendola:2000er,Bertolami:2000dp,Tocchini-Valentini:2001ty,Bean:2001zm}.
From the action Eq.
\eqref{equ::action}, we see that the mass of the fermions could be $\Phi$
dependent: $m_{\rm f} = \mf$. Two possible realization of this mass
dependence are for instance $m_{\rm f} = m^0_{\rm f} \exp(-\beta
\pmpc)$ and $m_{\rm f} = m^0_{\rm f} + c(\Phi_{\rm cl})$, where in the second
case, we may have a large field independent part together with small
couplings to quintessence.\footnote{The constant $\mfn$ is \emph{not} the fermion mass today, which
would rather be $m_{\rm today} = m_{\rm f}(\pmpc({\rm today}))$.}
 For the model discussed in
\cite{Amendola:2000er}, the coupling is of the first form, whereas in \cite{Bean:2001zm},
the coupling is realized by multiplying the cold dark
matter Lagrangian by a factor $f(\Phi$).
Hence, the coupling is $m_{\rm f}(\Phi) = f(\Phi)\, \mfn$, if we assume that
dark matter is fermionic. If it were bosonic, the following arguments would be similar.
We will only discuss the new effects coming from the coupling and
set $\vv{\eloop}{} = \vv{cl}{}  - \Delta V$,
where $\Delta V = \lferm^2 \mfsq / \left(8\pi^2\right)$.\\
Let us consider the case that all of the fermion mass is field dependent, i.e. we consider
cases like $m_{\textrm{f}}=m^{0}_{\textrm{f}}\exp(-\beta\pmpc)$.
The ratio of the `correction' to the classical potential is
\begin{equation}\label{overwhelming}
\frac{\Delta V}{\vv{cl}{}} = \frac{1}{8\pi^2} \frac{\lferm^2  \mfsq }{\vv{cl}{}}\approx 10^{80}.
\end{equation}
here we used a fermion cutoff at the GUT scale $\Lambda_{\textrm{ferm}}=10^{-3}$,
and a fermion mass of the order of $100 \textrm{Gev}=10^{-16}\mplank$ for an estimate.
Thus, the classical potential is negligible relative to the
correction induced by the fermion fluctuations.

Having made this estimate, it is clear that the fermion loop
corrections are only harmless, if the square of the coupling takes on
\emph{exactly} the same form as the classical potential itself. If,
for example we have an exponential potential
$\vv{cl}{} = A\exp(-\lambda \pmpc)$
together with a coupling $\mf = m^0_{\rm f}\exp(-\beta \pmpc)$, then
this coupling can only be tolerated, if $2\beta  = \lambda$.\footnote{Of course, sufficiently small $\beta$, will lead to a more
or less constant contribution, where $\mf \approx  m^0_{\rm f} -\beta \pmpc$.}
Taken at face value, this finding restricts models with these types of coupling.
It is however interesting to note that for the exponential coupling, the
case $2\beta = \lambda$ is not ruled out
by cosmological \mbox{observations \cite{Tocchini-Valentini:2001ty}.}\\
Turning to the possibility of a fermion mass that consists of
a field independent part and a coupling, i.e.
$m_{\rm f} = m^0_{\rm f} + c(\Phi_{\rm cl})$,
we find by similar reasoning that
\begin{equation}
\label{equ::cbound}
c(\Phi_{\rm cl})  \ll 3 \times 10^{-97},
\end{equation}
is needed in order for the potential to be stable.
Once again, the bound from Eq. \eqref{equ::cbound}
only applies if the functional form of the loop correction differs from the
classical potential.

The coupled models share one property: the loop contribution from the
coupling is by far larger than the classical potential. At first sight,
the golden way out of this seems to view the potential as already
effective: all fluctuations would be included from the start.
However, there is no particular reason, why \emph{any} coupling
of quintessence to dark matter should produce just exactly \emph{the}
effective potential used in a particular model: there is a relation between
a coupling and the effective potential generated.
Put another way,
\emph{if} the effective potential is of an elegant form and we have
a given coupling, then it seems unlikely that the \emph{classical}
potential could itself be elegant or natural.

Finally let us remark that the restrictions on a coupling to fermions are so severe
that we checked whether a gravitationally induced coupling could violate these bounds.
However the $c(\Phi)$-induced by graviton exchange is always proportional to $V(\Phi)$
and therefore the correction to the potential resulting from this effective coupling is
proportional to $V(\pmpc)$. Therefore, it does not change the form of the potential and
is harmless.
\section{Conclusions}
We have calculated quantum corrections to the classical
potentials of various quintessence models. In the late universe,
most potentials are stable with respect to the scalar quintessence
fluctuations.

An explicit coupling of the quintessence field to fermions (or similarly to dark matter
bosons) seems to be severely restricted. The effective potential to one loop level would be
completely dominated by the contribution from the fermion fluctuations.
All models in the literature share this fate. One way around this conclusion
could be to view these potentials as already effective. They must, however, not only
be effective in the sense  of an effective quantum field theory originating as a low-energy limit of
an underlying theory, but also include all fluctuations from this effective QFT.
In this case, there is a strong connection between coupling and potential and it
is rather unlikely that the \emph{correct} pair can be guessed.

The bound on the coupling is so severe that for consistency, we
have calculated an effective coupling due to graviton exchange.
To lowest order in $V(\Phi)$, this coupling leads to  a fermion contribution
which can be absorbed by redefining the pre-factor of the potential.

Surely, the one-loop calculation does not give the true effective potential.
Symmetries or more fundamental
theories that make the cosmological constant small as it is, could
force loop contributions to cancel. In addition, the back-reaction
of the changing effective potential on the fluctuations remains
unclear in the one loop calculation. A renormalization group
treatment would therefore be of great value. We leave this to
future work.
\section*{Acknowledgments}
We would like to thank Gert Aarts, Luca Amendola, J\"urgen Berges, Matthew Lilley,
Volker Schatz and Christof Wetterich for helpful discussions and the organizers for
a fruitful conference.

\end{document}